\documentclass[12pt]{article}%
\usepackage{amsmath,latexsym}
\usepackage{graphicx}
\usepackage{amsmath}
\usepackage{amsfonts}
\usepackage{amssymb}%
\begin{document}

\title{{A model for dark energy based on the theory
   of embedding}}
   \author{
Peter K.F. Kuhfittig* and Vance D. Gladney*\\
\footnote{kuhfitti@msoe.edu}
 \small Department of Mathematics, Milwaukee School of
Engineering,\\
\small Milwaukee, Wisconsin 53202-3109, USA}

\date{}
 \maketitle

\begin{abstract}\noindent
A long-standing topic of interest in the
general theory of relativity is the embedding
of curved spacetimes in higher-dimensional flat
spacetimes.  The main purpose of this paper is
to show that the embedding theory can account for
the accelerated expansion of the Universe and
thereby serve as a model for dark energy.  This
result is consistent with earlier findings
based on noncommutative geometry.  A secondary
objective is to show that the embedding theory
also implies that it is possible, at least in
principle, for the accelerated expansion to
reverse to become a deceleration.
\\
\end{abstract}

\noindent
\textbf{PAC numbers:} 04.20.Jb, 04.20.-q,
   04.50.+h\\


\section{Introduction}

Embedding theorems have a long history in the
general theory of relativity and continue to be
a topic of interest since they are able to
provide useful connections between the classical
theory and higher-dimensional spacetimes.  For
example, an $n$-dimensional Riemannian space is
said to be of embedding class $m$ if $n+m$ is
the lowest dimension of the flat space in which
the given space can be embedded.  In this paper
we begin with a generic line element of embedding
class two in a cosmological setting.  This line
element is reduced to class one by a suitable
transformation.  Our main objective is to show
that the resulting model is able to account for
the accelerated expansion of the Universe and
thereby serve as a model for dark energy.  It
is also shown that our conclusion is consistent
with earlier results based on noncommutative
geometry.

A secondary objective is to show that a change
in the nature of the embedding space could
result in a reversal of the accelerated expansion.

\section{The first embedding}\label{S:embedding}
\noindent
The discussion in Ref. \cite{MG17} begins with the
static and spherically symmetric line element
\begin{equation*}
ds^{2}=e^{\nu(r)}dt^{2}-e^{\lambda(r)}dr^{2}
-r^{2}\left(d\theta^{2}+\sin^{2}\theta \,d\phi^{2}
\right),
\end{equation*}
using units in which $c=G=1$.  It is shown in Ref.
\cite{MG17} that this metric of class two can be
reduced to a metric of class one by a suitable
transformation and can therefore be embedded in
the five-dimensional flat spacetime
\begin{equation*}
ds^{2}=-\left(dz^1\right)^2-\left(dz^2\right)^2
-\left(dz^3\right)^2-\left(dz^4\right)^2
+\left(dz^5\right)^2.
\end{equation*}

In our situation it is more convenient to employ the
opposite signature in the above line element:
\begin{equation}\label{E:line1}
ds^{2}=-e^{\nu(r)}dt^{2}+e^{\lambda(r)}dr^{2}
+r^{2}\left(d\theta^{2}+\sin^{2}\theta \,d\phi^{2}
\right).
\end{equation}
Similarly, the five-dimensional flat spacetime is
now written as follows:
\begin{equation}\label{E:line2}
ds^{2}=-\left(dz^1\right)^2+\left(dz^2\right)^2
+\left(dz^3\right)^2+\left(dz^4\right)^2
+\left(dz^5\right)^2.
\end{equation}
This reduction is accomplished by the following
transformation:
$z^1=\sqrt{K}\,e^{\frac{\nu}{2}}
 \,\text{sinh}{\frac{t}{\sqrt{K}}}$,
  $z^2=\sqrt{K}
 \,e^{\frac{\nu}{2}}\,\text{cosh}{\frac{t}{\sqrt{K}}}$,
 $z^3=r\,\text{sin}\,\theta\,\text{cos}\,\phi$, $z^4=
 r\,\text{sin}\,\theta\,\text{sin}\,\phi$,
 and
 $z^5=r\,\text{cos}\,\theta$.
The differentials of these components are

\begin{equation}
dz^1=\sqrt{K}\,e^{\frac{\nu}{2}}\,\frac{\nu'}{2}\,
\text{sinh}{\frac{t}{\sqrt{K}}}\,dr + e^{\frac{\nu}{2}}\,
\text{cosh}{\frac{t}{\sqrt{K}}}\,dt,
\end{equation}
\begin{equation}
dz^2=\sqrt{K}\,e^{\frac{\nu}{2}}\,\frac{\nu'}{2}\,
\text{cosh}{\frac{t}{\sqrt{K}}}\,dr + e^{\frac{\nu}{2}}\,
\text{sinh}{\frac{t}{\sqrt{K}}}\,dt,
\end{equation}
\begin{equation}
dz^3=dr\,\text{sin}\,\theta\,\text{cos}\,\phi + r\,
\text{cos}\,\theta\,\text{cos}\,\phi\,
d\theta\,-r\,\text{sin}\,\theta\,\text{sin}\,\phi\,d\phi,
\end{equation}

\begin{equation}
dz^4=dr\,\text{sin}\,\theta\,\text{sin}\,\phi + r\,
\text{cos}\,\theta\,\text{sin}\,\phi\,
d\theta\,+r\,\text{sin}\,\theta\,\text{cos}\,\phi\,d\phi,
\end{equation}
and
\begin{equation}
dz^5=dr\,\text{cos}\,\theta\, - r\,\text{sin}\,\theta\,d\theta,
\end{equation}
where the prime denotes differentiation with respect
to the radial coordinate $r$. To facilitate the substitution
into Eq. (\ref{E:line2}), we first obtain the expressions
for $-\left(dz^1\right)^2+\left(dz^2\right)^2$ and for
$\left(dz^3\right)^2+\left(dz^4\right)^2
+\left(dz^5\right)^2$:
\begin{equation}\label{E:partial1}
  -\left(dz^1\right)^2+\left(dz^2\right)^2=
  -e^{\nu}dt^{2}+\left(\,\frac{1}{4}K\,e^{\nu}\,
(\nu')^2\,\right)\,dr^{2}
\end{equation}
and
\begin{equation}\label{E:partial2}
  \left(dz^3\right)^2+\left(dz^4\right)^2
+\left(dz^5\right)^2=dr^2+r^{2}\left(d\theta^{2}
+\sin^{2}\theta\, d\phi^{2} \right).
\end{equation}
Substituting Eqs. (\ref{E:partial1}) and
(\ref{E:partial2}) into Eq. (\ref{E:line2}), we get
\begin{equation}\label{E:line3}
ds^{2}=-e^{\nu}dt^{2}+\left(\,1+\frac{1}{4}K\,e^{\nu}\,
(\nu')^2\,\right)\,dr^{2}+r^{2}\left(d\theta^{2}
+\sin^{2}\theta\, d\phi^{2} \right).
\end{equation}
Metric (\ref{E:line3}) is therefore equivalent to
metric (\ref{E:line1}) if
\begin{equation}\label{E:lambda1}
e^{\lambda}=1+\frac{1}{4}K\,e^{\nu}\,(\nu')^2,
\end{equation}
where $K>0$ is a free parameter.  The condition
is equivalent to the following condition due to
Karmarkar \cite{kK48}:
\[
   R_{1414}=\frac{R_{1212}R_{3434}+R_{1224}R_{1334}}
   {R_{2323}},\quad R_{2323}\neq 0.
\]
(See Ref. \cite{pB16} for further details.)  Other
useful references are
\cite{MM17, MRG17, sM17, MGRD, MDRK}.

\section{The problem}

It is well known that Alexander Friedmann
proposed in 1922 that our Universe cannot be
static, i.e., it must be either expanding or
contracting.  So if $a(t)$ is the scale factor
in the FLRW model, then one of the Friedmann
equations is
\begin{equation}\label{E:Friedmann}
  \frac{\overset{..}{a}(t)}{a(t)}=
  -\frac{4\pi}{3}(\rho +3p),
\end{equation}
again using units in which $c=G=1$.  Since we
are assuming that the entire Universe can be
embedded in a five-dimensional flat spacetime,
we necessarily find ourselves in a cosmological
setting.  So any point can be used as the
origin, since our Universe is a 3-sphere,
having neither a center nor an edge.  A
convenient way to proceed is to use the
following generic line element:
\begin{equation}\label{E:line4}
ds^{2}=-e^{\nu(r)}dt^{2}+e^{\lambda(r)}dr^{2}
+r^{2}\left(d\theta^{2}+\sin^{2}\theta \,d\phi^{2}
\right).
\end{equation}
We also need to assume that the spacetime is
asymptotically flat, i.e., $e^{\nu(r)}\
\rightarrow 1$ and $e^{\lambda(r)}
\rightarrow 1$ as $r\rightarrow\infty$.

Finally, let us list the first two of the
Einstein field equations \cite{pK17}:
\begin{equation}\label{E:E1}
   8\pi\rho=e^{-\lambda}\left(\frac
   {\lambda'}{r}-\frac{1}{r^2}\right)+
   \frac{1}{r^2}
\end{equation}
and
\begin{equation}\label{E:E2}
   8\pi p=e^{-\lambda}\left(\frac
   {1}{r^2}+\frac{\nu'}{r}\right)-
   \frac{1}{r^2}.
\end{equation}

Our immediate goal is to combine these
ideas to show that $\overset{..}{a}(t)$
in Eq. (\ref{E:Friedmann}) is positive.
That is the topic of the next section.

\section{The solution}

From Eq. (\ref{E:lambda1}) we obtain
\begin{equation*}
  \lambda'=\frac{1}{1+\frac{1}{4}K\,e^{\nu}\,(\nu')^2}
  \frac{1}{4}Ke^{\nu}[(\nu')^2+2\nu'\nu''].
\end{equation*}
Then it follows from Eqs. (\ref{E:E1}) and
(\ref{E:E2}) that
\begin{multline}\label{E:F1}
   8\pi (\rho +3p)=e^{-\lambda}\left[
   \frac{1/r}{1+\frac{1}{4}K\,e^{\nu}\,(\nu')^2}
   \frac{1}{4}Ke^{\nu}[(\nu')^3+2\nu'\nu'']
   -\frac{1}{r^2}\right]\\+\frac{1}{r^2}
   +3e^{-\lambda}\left(\frac{1}{r^2}+
   \frac{\nu'}{r}\right)-\frac{3}{r^2}.
\end{multline}
Finally, we observe that in view of Eq.
(\ref{E:lambda1}), Eq. (\ref{E:F1}) can also
be written as
\begin{multline}\label{E:F2}
   8\pi (\rho +3p)=-\frac{2}{r^2}+
   \frac{2e^{-\lambda}}{r^2}+e^{-\lambda}
   \left[\frac{1/r}{1+\frac{1}{4}K\,
   e^{\nu}\,(\nu')^2}\frac{1}{4}Ke^{\nu}
   [(\nu')^3+2\nu'\nu'']+\frac{3\nu'}{r}
   \right]\\=-\frac{2}{r^2}
      \frac{\frac{1}{4}Ke^{\nu}(\nu')^2}
      {1+\frac{1}{4}K\,e^{\nu}\,(\nu')^2}
   +\frac{1}{1+\frac{1}{4}K\,e^{\nu}\,(\nu')^2}
   \left[\frac{1/r}{1+\frac{1}{4}K\,
   e^{\nu}\,(\nu')^2}\frac{1}{4}Ke^{\nu}
   [(\nu')^3+2\nu'\nu'']+\frac{3\nu'}{r}
   \right].
\end{multline}
Given that $K$ is a free parameter, $K$ can
assume any positive value.  In fact, according
to Refs. \cite{MDRK, pK18}, $K$ can be extremely
large.  So for a sufficiently large $K$, the
second term on the right-hand side of Eq.
(\ref{E:F2}) becomes negligible, leading to
$8\pi(\rho +3p)<0$.  This shows that
$\overset{..}{a}(t)$ in the Friedmann
equation is positive, indicating an
accelerated expansion.

\section{The second embedding}

The idea of an extra spatial dimension had its
origin in the Kaluza-Klein theory and was
continued in the induced-matter theory in
Ref. \cite{WP92}.  One would normally assume
that the extra dimension has to be spacelike,
resulting in the signature $-++++$ in Eq.
(\ref{E:line2}).  It is proposed in Ref.
\cite{WP92}, however, that the signature
$--+++$ is in principle allowed, thereby
yielding two timelike components.  In other
words, the line element
\begin{equation*}
ds^{2}=-e^{2\nu_1(r)}dt_1^{2}-e^{2\nu_2(r)}dt_2^{2}
+e^{2\lambda(r)}dr^{2}
+r^{2}\left(d\theta^{2}+\sin^{2}\theta \,d\phi^{2}
\right)
\end{equation*}
would then be consistent with Einstein's theory.
To show that this is indeed the case, we note
that any calculation can be based on the
following orthonormal basis:  $\theta^0=
e^{\nu_1} dt_1$, $\theta^1=e^{\nu_2}dt_2$,
$\theta^2=e^{\lambda}\,dr$, $\theta^3=r\,dr$,
$\theta^4=r\,\text{sin}\,\theta\, d\phi$.
(See, for example, Ref. \cite{pK08}.)  Then
the line element takes on the form
\[
   ds^2=\mu_{ij}\theta^i\theta^j,
\]
where $\mu_{ij}$ is the usual Minkowski metric
$diag(-1,-1,1,1,1)$.  In other words, the
calculations are not affected by the signature.
So while ordinary $4D$ relativity does not
allow a second time coordinate, this is not
true in $5D$ since there is no conflict with
observation.

Returning to line element (\ref{E:line1}),
we can now consider an embedding in the
following five-dimensional flat spacetime:
\begin{equation*}
ds^{2}=-\left(dz^1\right)^2-\left(dz^2\right)^2
+\left(dz^3\right)^2+\left(dz^4\right)^2
+\left(dz^5\right)^2.
\end{equation*}
Then the transformation formulas for the first
two components must be changed to
$z^1=\sqrt{K}\,e^{\frac{\nu}{2}}
 \,\text{sin}{\frac{t}{\sqrt{K}}}$ and
  $z^2=\sqrt{K}\,e^{\frac{\nu}{2}}\,\text{cos}
  {\frac{t}{\sqrt{K}}}$;
  thus

\begin{equation*}
dz^1=\sqrt{K}\,e^{\frac{\nu}{2}}\,\frac{\nu'}{2}\,
\text{sin}{\frac{t}{\sqrt{K}}}\,dr + e^{\frac{\nu}{2}}\,
\text{cos}{\frac{t}{\sqrt{K}}}\,dt
\end{equation*}
and
\begin{equation*}
dz^2=\sqrt{K}\,e^{\frac{\nu}{2}}\,\frac{\nu'}{2}\,
\text{cos}{\frac{t}{\sqrt{K}}}\,dr - e^{\frac{\nu}{2}}\,
\text{sin}{\frac{t}{\sqrt{K}}}\,dt.
\end{equation*}
So
\[
   -\left(dz^1\right)^2-\left(dz^2\right)^2=
   -\frac{1}{4}Ke^{\nu}(\nu')^2\,dr^2-e^{\nu}\,dt^2.
\]
Since Eq. (\ref{E:partial2}) remains the same, the
resulting line element is
\begin{equation}\label{E:linenew}
ds^{2}=-e^{\nu}dt^{2}+\left(\,1-\frac{1}{4}K\,e^{\nu}\,
(\nu')^2\,\right)\,dr^{2}+r^{2}\left(d\theta^{2}
+\sin^{2}\theta\, d\phi^{2} \right)
\end{equation}
and
\begin{equation}\label{E:lambda2}
e^{\lambda}=1-\frac{1}{4}K\,e^{\nu}\,(\nu')^2,
\end{equation}

In view of line element (\ref{E:linenew}), $K$
cannot be arbitrarily large.  Now repeating the
earlier calculation, we arrive at
\begin{multline}
8\pi(\rho +3p)=\\\frac{2}{r^2}
      \frac{\frac{1}{4}Ke^{\nu}(\nu')^2}
      {1-\frac{1}{4}K\,e^{\nu}\,(\nu')^2}
   +\frac{1}{1-\frac{1}{4}K\,e^{\nu}\,(\nu')^2}
   \left[\frac{-1/r}{1-\frac{1}{4}K\,
   e^{\nu}\,(\nu')^2}\frac{1}{4}Ke^{\nu}
   [(\nu')^3+2\nu'\nu'']+\frac{3\nu'}{r}
   \right].
\end{multline}

To draw a conclusion, we need to make another
plausible assumption.  We know from the
asymptotic flatness that $\nu$ and $\nu'$ go
to zero as $r\rightarrow\infty$, but we also
assume that this occurs smoothly, as shown in
Fig. 1.   Now, with the Schwarzschild line
element in mind, let us assume that $\nu'>0$.
Then if $K$ is sufficiently small, we have
$8\pi(\rho+3p)>0$ and $\overset{..}{a}(t)<0$,
indicating a decelerating expansion.  Taken
together, these assumptions imply that a
reversal of the accelerated expansion is
indeed possible.

\section{Astronomy and the fifth dimension}

That the embedding theory in the first part
of this paper should yield an accelerated
expansion may not be a total surprise in
view of the induced-matter theory in Ref.
\cite{WP92}.  More precisely, it is noted
in Ref. \cite{pW15} that the field equations
for the five-dimensional flat embedding space
actually yield the Einstein field equations
in four dimensions \emph{containing matter}.
In other words, matter and hence energy,
including dark energy, come from geometry,
thereby echoing John A. Wheeler's
``everything is geometry."
Ref. \cite{pW13} goes on to state that even
the equivalence principle may be a direct
consequence of the existence of an extra
spatial dimension.  In fact, our very
understanding of physics in four dimensions
may be significantly enhanced due to this
extra dimension.

Another possible area of agreement is
noncommutative geometry, an offshoot of
string theory, which also involves extra
dimensions.  The main idea, discussed in
Refs. \cite{eW96, SW99}, is that
coordinates may become noncommuting operators
on a $D$-brane. Here the commutator is
$[\textbf{x}^{\mu},
\textbf{x}^{\nu}]=i\theta^{\mu\nu}$, where
$\theta^{\mu\nu}$ is an antisymmetric matrix.
The main idea, discussed in Ref. \cite{SS03},
is that noncommutativity replaces point-like
structures by smeared objects.  The
concentration is therefore on local properties.
It is shown in Ref. \cite{pK17b} that
$(-4\pi/3)(\rho+3p)>0$ locally, i.e., in
the neighborhood of every point.  This
suggests but does not necessarily prove
that the cumulative effect is an accelerated
expansion on a cosmological scale.  The
present study indicates, however, that this
interpretation is actually valid, given that
both theories depend on extra dimensions.
So noncommutative geometry is another way to
account for the accelerated expansion and
hence for dark energy.

The accelerated expansion is sometimes
referred to as a \emph{phase}.  The second
part of this paper explores the possibility of
just such a reversal due to a change in the
signature.  That such a change is possible
in the first place is shown by the
Schwarzschild line element: when crossing
the event horizon of a black hole, the first
two terms interchange signs.  While such a
change is not necessitated, it cannot be
excluded.  The same holds for the form of
$\nu(r)$.  So a reversal is indeed possible.

\emph{Remark:} Since mass and energy are
equivalent, dark energy generates a
gravitational field - one that is actually
repulsive.  If the above change in the
signature were to take place, thereby causing
the expansion to decelerate, then we may be
dealing with a more palatable kind of dark
energy - one that is gravitationally
attractive.

 \section{Conclusion}
An $n$-dimensional Riemannian space is said to
be of embedding class $m$ if $n+m$ is the
lowest dimension of the flat space in which the
given space can be embedded.  We start with a
spherically symmetric line element of embedding
class two which is then reduced to a metric of
class one by a suitable transformation.  This
metric may be viewed as a generic line element
in a cosmological setting.  The assumption of
asymptotic flatness then implies that $\nu(r)$
and $\nu'(r)$ in the line element go to zero
as $r\rightarrow\infty$.  An additional
assumption (needed only in the second part)
is that the asymptotic behavior occurs smoothly,
as shown qualitatively in Fig. 1.  The main
conclusion is that $(-4\pi/3)(\rho +3p)>0$;
so the Friedmann equation implies that
$\overset{..}{a}(t)>0$, where $a(t)$ is the
scale factor in the FLRW model.  The result
is an accelerated expansion of the Universe.
The embedding theory can therefore serve as
a model for dark energy.  The results are
consistent with earlier findings based on
noncommutative geometry.

The second part of the paper suggests that
a reversal of the accelerated expansion is
possible if the fifth dimension in the
embedding space becomes timelike and the
function $\nu(r)$ behaves in a manner
similar to that of the Schwarzschild line
element ($\nu(r)<0$, $\nu'(r)>0$).  While
the change in the signature is not
necessitated, it cannot be ruled out.
So it is possible, at least in principle,
for the acceleration to become a
deceleration.

\end{document}